\documentclass[aps,prb,showpacs,twocolumn]{revtex4}
\usepackage{bm,color,amsmath,amssymb,mathrsfs,latexsym,graphicx,psfrag}

\newcommand{\bs}[1]{\boldsymbol{#1}}

\newcommand{\bra}[1]{\left\langle#1\right|}
\newcommand{\ket}[1]{\left|#1\right\rangle}

\newcommand{\of}[1]{\!\left(#1\right)}

\newcommand{\expect}[1]{\left\langle#1\right\rangle}

\def\ie{{\it i.e.},\ }

\def\ea{{\it et al.}}
\def\eV{}

\begin{document}
\title{Numerical analysis of three-band models for CuO planes as
  candidates for a spontaneous T violating orbital current phase}
\author{Ronny Thomale and Martin Greiter}  
\affiliation{Institut f\"ur Theorie der Kondensierten Materie, 
  Universit\"at Karlsruhe, D 76128 Karlsruhe}
\pagestyle{plain}

%\homepage[]{Your web page}
%\thanks{}
%\altaffiliation{}
\date{\today}

% 0123456789012345678901234567890123456789012345678901234567890123456789
% 0         1         2         3         4         5         6
\begin{abstract}
%  We have numerically evaluated the current-current correlations for
%  three-band models of the CuO planes in high-$T_{\rm c}$
%  superconductors at hole doping $x=1/8$.  The results show no
%  evidence for the orbital current patterns proposed by Varma.  If
%  such patterns exist, the associated energy is estimated to be
%  smaller than 2.5 meV per link even if
%  $\epsilon_\text{p}-\epsilon_\text{d}=0$.  Assuming that the
%  three-band models are adequate, quantum critical fluctuations of
%  such patterns hence cannot be responsible for phenomena occurring at
%  significantly higher energies, such as superconductivity or the
%  anomalous properties of the pseudogap phase.
  Recently, we have numerically evaluated the current-current
  correlation function for the ground states of three-band models for
  the CuO planes of high-$T_{\rm c}$ superconductors at hole doping
  $x=1/8$ using systems with 24 sites and periodic boundary
  conditions.  In this article, the numerical analysis is explicated
  in detail and extended to a wider range of parameters.  Our results
  show no evidence for the time-reversal symmetry violating current
  patterns recently proposed by Varma.  If such current patterns
  exist, our results indicate that the energy associated with the loop
  currents must be smaller than 5 meV per link even if the on-site
  chemical potential on the oxygen sites, which is generally assumed
  to be of the order of 3.6 eV, is taken to zero, as advocated by
  Varma. We also vary the inter-atomic Coulomb repulsion scale and find
  only a weak dependence on this parameter. So while our studies do not
  rule out the existence of such current patterns, they do rule out
  that quantum critical fluctuations of these patterns are responsible
  for phenomena occurring at significantly higher energies such as the
  superconductivity or the anomalous properties observed in the
  strange metal phase provided the CuO superconductors are adequately
  described by any of the three-band models discussed.
\end{abstract}

\pacs{74.20.Mn, 74.72-h, 74.20.-z}
% 74.20.Mn  Nonconventional mechanisms (spin fluctuations, polarons 
%           and bipolarons, resonating valence bond model, anyon mechanism, 
%           marginal Fermi liquid, Luttinger liquid, etc.)
% 74.72.-h  Cuprate superconductors (high-Tc and insulating parent compounds)
% 74.20.-z  Theories and models of superconducting state
% insert suggested keywords - APS authors don't need to do this
%\keywords{}

\maketitle

\section{Introduction}
High-$T_{\rm c}$ superconductivity (HTSC) has been one of the most
active fields of research in condensed matter physics in the past two
decades~\cite{Bednorz-86zpb189, zaanen-06np138}. It has turned out to
be an exceedingly difficult problem, with much of the effort invested
just deepening the mysteries, but it has also led to a plethora of new
developments extending far beyond the field. Many ideas, even though
too general to qualify as complete theories of the cuprates, have
inspired a vast amount of research in both high-$T_{\rm c}$ and other
areas.  Most prominently among them are the notions of a resonating
valence bond (RVB) state~\cite{Anderson87s1196}, the gauge theories of
antiferromagnetism~\cite{Lee-06rmp17}, and the notion of quantum
criticality~\cite{Sachdev03rmp913}.  There have been, however, a few
concise proposals which make falsifiable predictions.  Intellectual
masterpieces among them have been the theory of anyon
superconductivity~\cite{Laughlin88s525}, the proposal of kinetic
energy savings through interlayer tunneling~\cite{Anderson95s1154},
the SO(5) theory of a common order parameter for superconductivity and
magnetism~\cite{Demler-04rmp909}, and a more recent proposal that the
anomalous properties of the cuprates may be due to quantum critical
fluctuations of current patterns formed spontaneously in the CuO
planes~\cite{Varma99prl3538, Varma06prb155113}. In a recent
Letter~\cite{Greiter-07prl027005}, we investigated this proposal by finite
cluster calculations. %In this article, 
Here, we provide supplemental
information and a more elaborate account of the
approach.  

This paper is organized as follows. In Section~\ref{sec:prop}, we
briefly review Varma's proposal of spontaneous T violation in the CuO
planes and discuss some assumptions made therein.  In
Section~\ref{sec:mod}, we introduce the three-band model Hamiltonian
we investigate.  In Section~\ref{sec:cc}, we compute the
current-current correlations of the ground state which we use to
obtain information about the existence of orbital currents and the
magnetic moment associated with them.  We find that if an orbital current
phase exists in the cuprates, the energy associated with the
spontaneous currents will not be sufficiently high for the phase to
account for the strange metal phenomenology in the cuprates.  We
derive an upper bound for the magnetic moment per unit cell from the
upper bound we obtain for spontaneous currents, and find it smaller
than the magnetic moment measured in a recent neutron scattering
experiment~\cite{Fauque-06prl197001}.  The comparison shows, however,
that even if the observed magnetic moments were due a current pattern
as proposed by Varma~\cite{Varma99prl3538, Varma06prb155113}, the
magnitude of these currents would be insufficient to determine the
phase diagram.  In search for an alternative explanation of the
experiment, we investigate the spin-spin correlations of the ground
state in Section~\ref{sec:ss}. In Section~\ref{sec:num}, the results
for various three-band model parameters are presented. In particular,
we vary the on-site chemical potential on the oxygens
$\epsilon_{\text{p}}$ from previously 3.6 to 1.8, 0.9, 0.4, and
finally 0 eV.  Furthermore, we vary the Coulomb interaction scale
$V_{\text{pd}}$ from $1.2$ to $2.4$ eV.  The conclusion regarding the
relevance of an orbital current pattern for the strange metal phase of
CuO superconductors we reached previously remain intact.  In
Section~\ref{sec:finite}, we discuss the the role of finite size
effects in our numerical experiments, with particular emphasis on the
net spin 1/2 of our finite size ground states.  In
Section~\ref{sec:con}, we conclude that while we cannot rule out that
the orbital current phase exists in the cuprates, we can infer that
the energy associated with these fluctuations is not sufficiently high
to account for the strange metal phase in the cuprates.

\section{Hypothesis of spontaneous T violation in the cuprates}
\label{sec:prop}
The proposal of a spontaneous symmetry breaking through orbital
currents is motivated by experiment.  The normal state of the cuprates
at optimal doping shows a behavior which can be classified as quantum
critical, and has been rather adequately described by a
phenomenological theory called marginal Fermi
liquid~\cite{Varma-89prl1996}.  The linear temperature dependence of
the normal-state resistivity in optimally doped LSCO and
YBCO~\cite{Gurvitch-87prl1337, Takagi-92prl2975,dagan-04prl167001},
which persists over several magnitudes of temperature, provides
striking evidence in favor of this picture.
%
%(Recent experiments by \ldots in ultra-pure \ldots samples show a
%negative offset for the resistivity if the straight line is
%extrapolated to $T=0$, which precludes an interpretation in terms of
%quantum critical fluctuations, but these data where not available when
%the theory was proposed.)
%
The marginal Fermi liquid phenomenology led Varma to assume 
%This phenomenology suggests 
a quantum critical point (QCP) at a hole
doping level of $x_{\rm c}\approx 0.19$, an assumption %which appears
consistent with a significant body of experimental
data~\cite{Tallon-01pc53, Alff-03n698, vanderMarel03n271,
dagan-04prl167001, Naqib-05prb054502}.  Critical fluctuations around
this point are then held responsible for the anomalous properties of
the strange metal phase, and provide the pairing force responsible for the
superconducting phase which hides the QCP.

Interpreting the phase diagram in these terms, one is immediately led
to ask what the phase to the left of the QCP, \ie for $x<x_{\rm c}$,
might be.  The theory would require a spontaneously broken symmetry
beyond the global U(1) symmetry broken through superconductivity.
%(which is often erroneously refered to as a broken gauge
%symmetry~\cite{Greiter05ap217})
In addition, as the fluctuations are
assumed to determine the phase diagram up to temperatures of several
hundred Kelvin, the characteristic energy scale of the correlations
associated with this symmetry violation must be at least of the same order of
magnitude.  No definitive evidence of such a broken symmetry has been
found up to now, even though several possibilities have been suggested.
These include stripes~\cite{Kivelson-03rmp1201}, a
$d$-density wave~\cite{Chakravarty-01prb094503}, and a
checkerboard charge density wave~\cite{Li-06prb184515}.

The general consensus is that the low energy sector of the three-band
Hubbard model proposed for the CuO planes~\cite{emery87prl2794} (see
\eqref{3bH} below) reduces to a one-band $t$--$t'$--$J$ model, with
parameters $t\approx 0.44$, $t'\approx 0.06$, and $J\approx 0.128$
(energies throughout this article are in eV)~\cite{Zhang-88prb3759,
  Eskes-88prl1415, Hybertsen-89prb9028, Hybertsen-90prb11068,
  Rice-91ptrsla459}.  However, two remarks are in order. First, the
parameters are not exactly known, but can only be calculated
approximately~\cite{Belinicher-96prb335}.  Second, for certain regimes
of the phase diagram, CuO two-leg ladder studies have shown that the
one-band and three-band description lead to qualitatively different
results~\cite{Lee-05prb075126,Fjaerestad-06ap894}. 

For the undoped CuO planes, the formal valances are Cu$^{2+}$ and
O$^{2-}$.  As the electron configuration of Cu atoms is [Ar]
$3d^{10}4s^1$, this implies one hole per unit cell, which will
predominantly occupy the $3d_{x^2-y^2}$ orbital.  As the on-site
potential $\epsilon_\text{p}$ in the O $2p_x$ and $2p_y$ orbitals
relative to the Cu $3d_{x^2-y^2}$ orbital is generally assumed to be
of the order of $\epsilon_\text{p}=3.6\eV$ (with
$\epsilon_{\text{d}}=0$), and hence smaller than the on-site Coulomb
repulsion $U_{\text{d}}\approx 10.5\eV$ for a second hole in the
$3d_{x^2-y^2}$ orbital, additional holes doped into the planes will
primarily reside on the Oxygens.  The maximal gain in hybridization
energy is achieved by placing the additional hole in a combination of
the surrounding O $2p_x$ and $2p_y$ orbitals with the same symmetry as
the original hole in the Cu $3d_{x^2-y^2}$ orbital, which requires
antisymmetry of the wave function in spin space, \ie the two holes
must form a singlet.  This picture is strongly supported by data from
NMR~\cite{Walstedt90s248} and even more directly from spin-resolved
photoemission~\cite{tjeng-97prl1126}.  In the effective one-band
$t$--$J$ model description of the CuO planes, these singlets
constitute the ``holes'' moving in a background of spin 1/2 particles
localized at the Cu sites.
\begin{figure}[t]
% \setlength{\unitlength}{2pt}
% \begin{picture}(100,60)(0,0)
% \multiput(20,57.7)(20,0){2}{\makebox(0,0){\small O}}
% \multiput(0,37.7)(20,0) {4}{\makebox(0,0){\small O}}
% \multiput(0,17.7)(20,0) {4}{\makebox(0,0){\small O}}
% \multiput(20,-2.3)(20,0) {2}{\makebox(0,0){\small O}}
% \multiput(30,47.7)(40,0){1}{\makebox(0,0){\small Cu}}
% \multiput(10,27.7)(40,0){2}{\makebox(0,0){\small Cu}}
% \multiput(30,7.7)(40,0){1}{\makebox(0,0){\small Cu}}
% \multiput(30,60)(20,0){1}{\makebox(0,0){\vector(1,0){14}}}
% \multiput(10,40)(20,0){3}{\makebox(0,0){\vector(1,0){14}}}
% \multiput(10,20)(20,0){3}{\makebox(0,0){\vector(1,0){14}}}
% \multiput(30,0)(20,0){1}{\makebox(0,0){\vector(1,0){14}}}
% \multiput(27,50)(10,-10){4}{{\vector(-1,1){5}}}
% \multiput(7,30)(10,-10){4}{{\vector(-1,1){5}}}
% \multiput(37,55)(-10,-10){4}{{\vector(-1,-1){5}}}
% \multiput(57,35)(-10,-10){4}{{\vector(-1,-1){5}}}
% \end{picture}
  \begin{minipage}[c]{0.4\linewidth}
    \includegraphics[width=\linewidth]{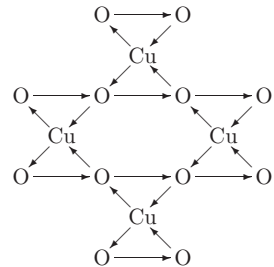}
  \end{minipage}
\caption{Orbital current pattern proposed by Varma.}
\label{fig:pattern}
\vspace{-4mm}
\end{figure}

In contrast to this picture, Varma~\cite{Varma99prl3538,
  Varma06prb155113} has proposed that the additional holes doped in
the CuO planes do not hybridize into Zhang-Rice singlets, but give
rise to circular currents on O-Cu-O triangles, which align into a
planar pattern as shown in Fig.~\ref{fig:pattern}.  He assumes that
the inter-atomic Coulomb potential $V_\text{pd}$ is larger than both
the hopping $t_\text{pd}$ and the on-site potential $\epsilon_\text{p}$
of the O $2p$ orbitals relative to the Cu $3d_{x^2-y^2}$ orbitals, an
assumption which is not consistent with the values generally agreed on
(see the list below \eqref{3bH}).
% proposed by Hybertsen \ea, who assumed $V_\text{pd}=1.2\eV$,
% $t_\text{pd}=1.5\eV$, and $\epsilon_\text{p}=3.6\eV$.
Making additional assumptions, Varma has shown that the circular
current patterns are stabilized in a mean field solution of the
three-band Hubbard model.  The orbital current patterns break
time-reversal symmetry (T) and the discrete four-fold
rotation symmetry on the lattice, but leave translational symmetry
intact.  The current pattern is assumed to disappear at a doping level
of about $x_{\rm c}\approx 0.19$.  The phenomenology of CuO
superconductors, including the pseudogap and the marginal Fermi liquid
phase, are assumed to result from critical fluctuations around this
QCP, as outlined above.

Motivated by this proposal, several experimental groups have looked
for signatures of orbital currents or T violation in CuO
superconductors.  While there is no agreement between different groups
regarding the manifestation of T violation in ARPES 
studies~\cite{Kaminski-02n610,Borisenko-04prl207001}, a recent
neutron scattering experiment by Fauqu\'e \ea~\cite{Fauque-06prl197001} 
%carried out in P.\ Bourges' group
indicates magnetic order within the unit cells of the CuO planes.
Their results appear to be consistent with Varma's proposal, and call the
validity of the one-band models into question.  
%This experiment has stimulated intense interest in Varma's proposal.
In a recent article, Aji and Varma~\cite{Aji-06prl067003} have mapped
the four possible directions of the current patterns in each unit cell
onto two Ising spins, and investigated the critical fluctuations.
Within this framework, the couplings between and the transverse fields
for these Ising spins effect whether or under which circumstances the
model displays long-range order in the orbital currents.  

We hence intended to estimate these couplings through numerical
studies of finite clusters containing 8 unit cells, \ie 8 Cu and 16 O
sites, and periodic boundary conditions (which do not frustrate but
should enhance the correlations).  The total number of holes on our
cluster was taken $N=9$ (5 up-spins and 4 down-spins), corresponding
to a hole doping of $x=1/8$.  We had hoped that the energy associated
with a domain wall, which may be implemented through a twist in the
boundary conditions, would provide information regarding the coupling
aligning the orbital currents in neighboring plaquets, while the
splitting between the lowest energies for a finite system would
provide an estimate for the transverse field. Together, this would
account for a description of the system in terms of Ising-type
variables where the quantum critical behavior could be analyzed.

We find, however, that the current-current correlations in the ground
state show no tendency to align the orbital currents whatsoever.  There
is not even a context to speak of a coupling of these Ising
variables---or, in other words, the couplings are zero within the
error bars of our numerical experiments.
\section{Three-band model for the cuprates}
\label{sec:mod}
To begin with, we wish to study the three-band Hubbard Hamiltonian
$H=H_\text{t}+H_\text{U}$ with~\cite{Dagotto94rmp763}
\begin{eqnarray}
  H_\text{t}&\!=\!&\! 
  \sum_{i,\sigma} \epsilon_\text{p}\, 
  n_{i,\sigma}^{\text{p}}
  -t_{\text{pd}}\!\sum_{\langle i,j \rangle , \sigma}\! 
  \left(d_{i,\sigma}^{\dagger} p^{\phantom{\dagger}}_{j,\sigma}
    +p_{j,\sigma}^{\dagger}d_{i,\sigma}^{\phantom{\dagger}}\right)
  \nonumber \\
  &-\!&\! t_{\text{pp}}\!\sum_{\langle i,j \rangle , \sigma}\!
  \left( p_{i,\sigma}^{\dagger}p_{j,\sigma}^{\phantom{\dagger}}
    +p_{j,\sigma}^{\dagger}p_{i,\sigma}^{\phantom{\dagger}}\right)
  + V_{\text{pd}}\hspace{-9pt}\sum_{\langle i,j \rangle , \sigma, \sigma '}
  \hspace{-9pt} n_{i, \sigma}^{\text{d}} n_{j,\sigma '}^{\text{p}},\hspace{3pt}  
  \nonumber \\[2pt]
  H_\text{U}\!\!&\!=\!&\! 
  U_{\text{p}}\sum_i n_{i, \uparrow}^{\text{p}} n_{i, \downarrow}^{\text{p}}
  + U_{\text{d}}\sum_i n_{i, \uparrow}^{\text{d}} n_{i, \downarrow}^{\text{d}},
  \label{3bH}
\end{eqnarray} 
where {\small $\langle\ ,\ \rangle$} indicates that the sums extend
over pairs of nearest neighbors, while $d_{i,\sigma}$ and
$p_{j,\sigma}$ annihilate holes in Cu $3d_{x^2-y^2}$ or O $2p$
orbitals, respectively.  Hybertsen \ea~\cite{Hybertsen-89prb9028}
calculated $t_{\text{pd}}=1.5\eV$, $t_{\text{pp}}=0.65\eV$,
$U_{\text{d}}=10.5\eV$, $U_{\text{p}}=4\eV$, $V_{\text{pd}}=1.2\eV$,
and $\epsilon_{\text{p}}=3.6\eV$, which is the first three-band model
discussed below. 
% Note that whereas the sign
%for $t_{\text{pd}}$ is irrelevant and can be trivially removed by a
%gauge transformation, it is important to note that in hole notation,
%the orbital hopping from a $p_{\text{x}}$ to a $p_{\text{y}}$ Oxygen
%orbital enters with $-t_{\text{pp}}$. 

%\begin{figure}[t]
%  \begin{minipage}[c]{0.9\linewidth}
%    \includegraphics[width=\linewidth]{paper-ep36}
%%    \includegraphics[width=\linewidth]{sq}
%  \end{minipage}
%  \caption{Current-current correlations $\expect{j_{k,k+\hat
%        x}j_{l,m}}$ multiplied by $10^2$ for the ground state of
%    \eqref{3btj} with $\epsilon_\text{p}=3.6$ on a 24 site cluster (8
%    Cu = open circles, 16 O = filled circles) with PBCs.  The
%    reference link is indicated in the top and (due to the PBCs)
%    bottom left corner.}
%\label{fig:ep36}
%\vspace{-3mm}
%\end{figure} 

In order to be able to diagonalize \eqref{3bH} for a cluster of 24
sites, \ie 8 Cu and 16 O sites, with 5 up-spin and 4
down-spin holes, we need to truncate the Hilbert space.
A first step is to eliminate doubly occupied sites. This yields the
effective three-band $t$--$J$ Hamiltonian
\begin{eqnarray}
  H_{\text{eff}}\!\!&=\!\!&\tilde{P}_{\text{G}}  H_\text{t}
  \tilde{P}_{\text{G}} + H_\text{J} \ \ \ \text{with}\nonumber\\[5pt]
  H_\text{J}&\!\!=\!\!&
  J_{\text{pd}}\hspace{-1pt}\sum_{\langle i, j \rangle} 
  \of{\bs{S}_{i}^{\text{p}}\!\cdot\! \bs{S}_{j}^{\text{d}}-\frac{1}{4}}
  + J_{\text{pp}}\hspace{-1pt}\sum_{\langle i, j \rangle} 
  \of{\bs{S}_{i}^{\text{p}}\!\cdot\! \bs{S}_{j}^{\text{p}}-\frac{1}{4}},
  \hspace{5pt}\nonumber\\
  \label{3btj}
\vspace{-5pt}
\end{eqnarray}
where
% $J_{\text{pd}}=2t_{\text{pd}}^2
% \of{\frac{1}{U_{\text{d}}-\epsilon_\text{p}}
% +\frac{1}{U_\text{p}+\epsilon_\text{p}}}=1.55\eV$
% and $J_{\text{pp}}=4t_{\text{pd}}^2/U_{\text{p}}=0.42\eV$.  
\begin{equation}
  \label{jpdjpp}
  J_{\text{pd}}=2t_{\text{pd}}^2\of{\frac{1}{U_{\text{d}}-\epsilon_\text{p}}
    +\frac{1}{U_\text{p}+\epsilon_\text{p}}}, \
  J_{\text{pp}}=\frac{4t_\text{pp}^2}{U_\text{p}}\hspace{1pt},
  \nonumber
\end{equation}
and the sums in $H_\text{J}$ are limited to pairs where both neighbors
are occupied by holes.  If $\tilde{P}_{\text{G}}$ only eliminates
configurations with more than one hole on a site, \ie a pure
Gutzwiller projection, the dimension of the
$S^z_\text{tot}=\frac{1}{2}$ sub-sector is 164,745,504, which as such
is above our capabilities.

\begin{figure}[t]
  \begin{minipage}[c]{0.9\linewidth}
    \includegraphics[width=\linewidth]{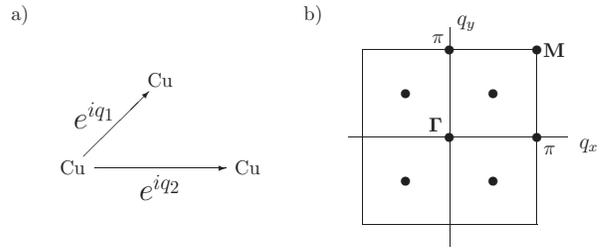}
  \end{minipage}
  \caption{We define the phases $q_1$ and $q_2$ acquired by
    translations shown in a). For a system of eight unit cells, \ie 8
    Cu and correspondingly 16 O, the phases can have the values
    $q_1=n_1\pi/2$ and $q_2=n_2 \pi$ with $n_1\in \{0,1,2,3\}$ and
    $n_2\in \{0,1\}$. For the Brillouin zone depicted in b), $q_1$ and
    $q_2$ translate into $x$ and $y$ momenta by $q_x=q_1$ and
    $q_y=q_1-q_2$. Thus, the phase description $(q_1,q_2)$ of the
    $\Gamma$ and M point is $(0,0)$ and $(\pi,0)$, respectively.}
\label{fig:bz}
\vspace{-4mm}
\end{figure}
As a next step, we exploit the translational symmetries on the
cluster, a 4-fold symmetry in $\hat{q_1}$ direction and a 2-fold
symmetry in $\hat{q_2}$ direction according to the conventions given
in Fig.~\ref{fig:bz}. (Throughout this article, we label the momenta
by $(q_1,q_2)$ rather than $(q_x,q_y)$.) The reduced dimension $\sim 2
\cdot 10^{8}$ is still a considerably large Hilbert space.  Thus, we
take two further steps. First, we identify the ground state in the
Brillouin zone. As it turns out, for all parameter choices discussed
in this article, the respective ground state is either at the $\Gamma$
or M point. Second, we apply the rotation symmetry, which commutes
with the translational symmetries at and only at the M and $\Gamma$
point. The system then becomes amenable to exact diagonalization.

\begin{table}[b]
  \centering
  \begin{tabular}{ccccc}
    \hline\hline
    $N^\text{max}_\text{ox}$   & (0,0) & ($\frac{\pi}{2}$,0) 
    & ($\pi$,0) & (0,$\pi$)  \\
    \hline
    3& -0.7025  & -0.6948  & -0.7051  & -0.6924    \\
    4& -0.8256  & -0.8204  & -0.8350  & -0.8198    \\
    5& -0.8719  & -0.8611  & -0.8774  & -0.8617    \\
    \hline\hline
  \end{tabular}
  \caption{Ground state energies per unit cell for 
    $\epsilon_{\text{p}}=3.6$, $V_{\text{pd}}=1.2$ for the inequivalent 
    points in the Brillouin zone and truncation (a).}
  \label{tab:azero}
\end{table}
\begin{table}[b]
  \centering
  \begin{tabular}{ccccc}
    \hline\hline
    $N^\text{max}_\text{link}$   & (0,0) & ($\frac{\pi}{2}$,0) 
    & ($\pi$,0) & (0,$\pi$) \\
    \hline
    2& \phantom{-}0.0640  & \phantom{-}0.0693  & \phantom{-}0.0631  
    & \phantom{-}0.0715  \\
    3& -0.6285  & -0.6224  & -0.6351  & -0.6215  \\
    4& -0.8473  & -0.8356  & -0.8520  & -0.8359  \\
    \hline\hline
  \end{tabular}
  \caption{Ground state energies per unit cell for 
    $\epsilon_{\text{p}}=3.6, \; V_{\text{pd}}=1.2$ for the inequivalent 
    points in the Brillouin zone and truncation (b).}
  \label{tab:bzero}
\end{table}
In order to identify the momentum of the ground state, we introduce
two ways of truncating the Hilbert space: (a) We limit the maximal
number of holes allowed in the O orbitals to $N^\text{max}_\text{ox}$.
(b) We limit the maximal number of CuO links occupied with 2 holes to
$N^\text{max}_\text{link}$.
Truncation (a) serves as a good scheme when the on-site potential
$\epsilon_{\text{p}}$ is large compared to other parameter scales, but
is not practicable in other cases.  Since truncation (b) predominantly
projects out states with high kinetic energy, we expect it to be
insensitive to the value of $\epsilon_{\text{p}}$. To check the
validity of the truncations, we consider Hamiltonian~\eqref{3btj} with
the parameter values by Hybertsen {\it et al.} listed above and
calculate the ground state energies of the system (see Table~\ref{tab:azero}
and~\ref{tab:bzero}).

Both truncation schemes yield similar results. The ground state is
situated at the M point $(\pi,0)$ of the Brillouin zone. At this
point, it is possible to implement a 4-fold rotation symmetry, which
commutes with the translational symmetries. Thus, the dimension is
reduced by an additional factor of 4 and the exact state is
accessible. We find the energies $E_{(\pi,0,0)}=-0.8513$,
$E_{(\pi,0,\pi/2)}=-0.8570$, and $E_{(\pi,0,\pi)}=-0.8883$, where the
first two indices label the linear momenta $(q_1,q_2)$, and the third
labels the angular momentum under discrete rotations by $90^{\circ}$,
$q_{\text{rot}}=n_{\text{rot}}\pi / 4$, with
$n_{\text{rot}}\in\{0,1,2,3\}$. The ground state is hence in the
$(\pi,0,\pi)$ subspace.

\begin{figure}[t]
  \begin{minipage}[c]{0.9\linewidth}
    \includegraphics[width=\linewidth]{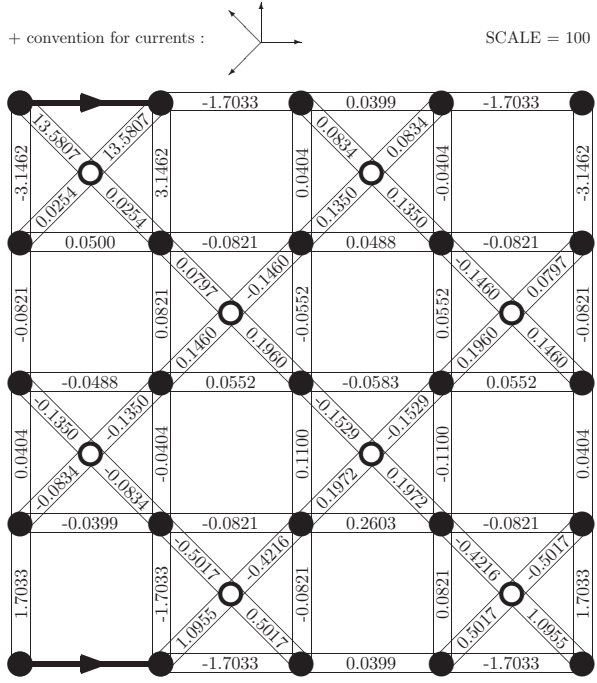}
  \end{minipage}
  \caption{Current-current correlations $\expect{j_{k,k+\hat
        x}j_{l,m}}$ multiplied by $10^2$ and in units of
    $\frac{eV}{\hbar}$ for the ground state of \eqref{3btj} with
    $\epsilon_\text{p}=3.6$, $V_{\text{pd}}=1.2$ on a 24 site cluster
    (8 Cu = open circles, 16 O = filled circles) with PBCs.  The
    reference link is indicated in the top and (due to the PBCs)
    bottom left corner. Except for the vertical lines, positive
    numbers indicate alignment with the pattern shown in
    Fig.~\ref{fig:pattern}.}
\label{fig:ep36}
%\vspace{-3mm}
\end{figure} 

\begin{figure}[t]
  \begin{minipage}[c]{0.9\linewidth}
    \includegraphics[width=\linewidth]{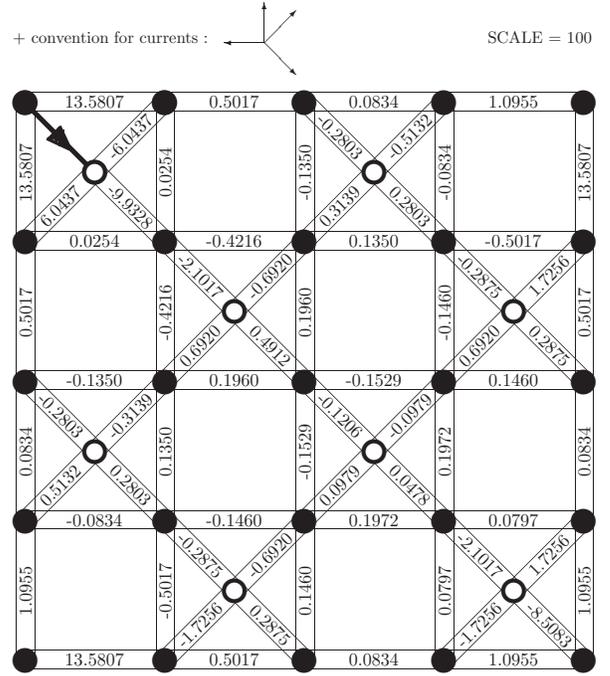}
  \end{minipage}
  \caption{Current-current correlations $\expect{j_{k,k+\hat
        x}j_{l,m}}$ multiplied by $10^2$ for the ground state of
    \eqref{3btj} with $\epsilon_\text{p}=3.6$, $V_{\text{pd}}=1.2$ on
    a 24 site cluster (8 Cu = open circles, 16 O = filled circles)
    with PBCs.  The Cu-O reference link is indicated in the top
    corner.}
\label{fig:ep36-2}
%\vspace{-3mm}
\end{figure}

\section{Current-current correlations and magnetic moments}
\label{sec:cc}

With the current operator for an O-O and a Cu-O link given by
\begin{equation}
  \label{curdef}
  j_{k,l}=\frac{i t_{\text{pp}}}{\hbar}\sum_\sigma
  \of{p_{l,\sigma}^{\dagger}p_{k,\sigma}^{\phantom{\dagger}}
     -p_{k,\sigma}^{\dagger}p_{l,\sigma}^{\phantom{\dagger}}}
\vspace{-3pt}
\end{equation}
and
\begin{equation}
  \label{curdefcuo}
  j_{k,l}=\frac{i t_{\text{pd}}}{\hbar}\sum_\sigma
  \of{p_{l,\sigma}^{\dagger}d_{k,\sigma}^{\phantom{\dagger}}
     -d_{k,\sigma}^{\dagger}p_{l,\sigma}^{\phantom{\dagger}}},
\vspace{-3pt}
\end{equation}
respectively,
we evaluated the correlation function $\expect{j_{k,k+\hat x}j_{l,m}}$
with O-O and Cu-O links as reference links for the exact ground state. 
The results are shown in Figs.~\ref{fig:ep36} and \ref{fig:ep36-2}. The
correlations fall off rapidly and there is no indication of order.

We now use the correlations to reach a conclusion regarding the
existence or non-existence of the orbital current pattern, indicated
in Fig.~\ref{fig:pattern}. The numerical experiments for the finite
cluster can, as a matter of principle, never rule out directly that a
symmetry, in our case time reversal symmetry T, is violated.  As only
real parameters enter the Hamiltonian, the computed ground states are
real by construction and do not allow for a direct indication of time
reversal symmetry breaking. If it were to exist, the computed ground
state would be a symmetric superposition of the different separately T
violating ground states, which itself is T symmetric again and
described by a real wave function. The current-current correlation
function, however, allows to put an upper bound on the size of the
spontaneous currents: If a current pattern as sketched in
Fig.~\ref{fig:pattern} were to exist, the current-current correlations
$\expect{j_{k,k+\hat x}j_{l,l+\hat x}}$ for links far away from each
other in a rotationally invariant ground state should approach
$\frac{1}{2}\expect{\hat x|j_{k,k+\hat x}|\hat x}^2$, where $\ket{\hat
  x}$ denotes a state with a spontaneous current pointing in $\hat x$
direction (the factor $\frac{1}{2}$ arises because by choosing our
reference link in $x$-direction, we effectively project onto two of
the four possible directions for the current pattern).  From the
values of $10^2\!\cdot\!\expect{j_{k,k+\hat x}j_{l,l+\hat x}}$ for the
four horizontally connected links in the center of
Fig.~\ref{fig:ep36}, $-0.0488, +0.0552, -0.0583$, and $+0.0552$, which
should all be positive if a current pattern were present, we estimate
$10^2\!\cdot\!\expect{j_{k,k+\hat x}}^2 < 0.05$ and hence
$\expect{\hat x|j_{k,k+\hat x}|\hat x}^2 < 1\cdot 10^{-3}$ as an upper
bound for a current pattern we are unable to detect through the
%``noise''.  
error bars of our numerical experiment. (Throughout this article,
currents are quoted in units of $\rm{eV} / \hbar$.)  We now denote
$\expect{\hat x|j_{k,k+\hat x}|\hat x}$ by
$j_\text{pp}$. 

We roughly estimate the kinetic energy $\varepsilon_\text{pp}$ per
link associated with a spontaneous current $j_\text{pp}$ of this
magnitude using $j_\text{pp}=n_\text{p}v$ and
$\varepsilon_\text{pp}=\frac{1}{2}n_\text{p} m v^2$ with
$m={1}/{2t_\text{pp}}$, where $n_\text{p}$ is the hole density on the
Oxygen sites ($n_\text{p}=0.14$ for the state analyzed in
Fig.~\ref{fig:ep36}), and obtain
\begin{equation}\label{eq:epp}
  \varepsilon_\text{pp}
  \approx\frac{\hbar^2 j_\text{pp}^2}{4t_\text{pp}n_\text{p}}< 3 \cdot 10^{-3}.
\end{equation}

A similar analysis with a Cu-O reference link (shown in
Fig.~\ref{fig:ep36-2}) yields with $10^2\!\cdot\!\expect{j_{k,k+\hat
    x+\hat y}}^2 < 1.0$ and hence $j_\text{pd}^2 < 10^{-2}$ (there is
no factor $\frac{1}{2}$ in this case) an estimate of
\begin{equation}\label{eq:epd}
%  \label{estimate-pd}
  \varepsilon_\text{pd}
  \approx\frac{\hbar^2 j_\text{pd}^2}{4t_\text{pd}\sqrt{n_\text{p}n_\text{d}}}
  <5 \cdot 10^{-3},
\end{equation}
where we have determined $n_\text{d}$ via $8n_\text{d}+16n_\text{p}=9$.

Note that this energy of 3 (or 5) meV is
not the condensation energy $E_{\text{c}}$ per unit cell, but a
positive contribution to the energy of the current carrying state,
which would have to be (more than) offset by other contributions (like
the energy gain from aligning the circulating currents according to
the pattern Varma proposed) if such a state were realized. We would
expect the transition temperature $T_{\text{c}}$ of such a state to be
of the order of the effective coupling of the Ising spins introduced
by Aji and Varma~\cite{Aji-06prl067003}, while we would expect that
$E_{\text{c}}\ll T_{\text{c}}$. 

Making contact to the quantities observed in experiment, we now
derive an upper bound of the magnetic moment from the upper bound
for the spontaneous currents we obtained through numerics. The
magnetization is related to the angular momentum of circulating
electrons by $M=-\mu_{\text{B}} L_z / \hbar$ with $L_z=m_e r v$, where
$r$ denotes the distance to the center of rotation and $v$ the
velocity of the electrons.

\begin{figure}[h]
  \begin{minipage}[c]{0.5\linewidth}
    \psfrag{A}{$\frac{a_0} {2}$}
    \psfrag{B}{$\frac{a_0}{6}$}
    \psfrag{C}{$r_{\text{pp}}$}
    \psfrag{D}{$j_{\text{pp}}$}
    \psfrag{E}{$r_{\text{pp}}=a_0 / 6\sqrt{2}$}
    \psfrag{F}{$r_{\text{pd}}$}
\includegraphics[width=\linewidth]{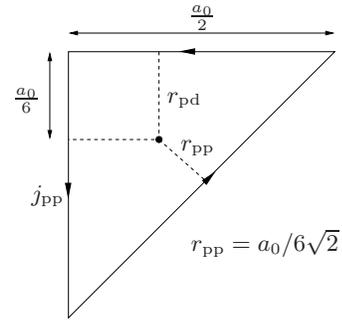}
  \end{minipage}
  \caption{Circulating current $j_{\text{pp}}$ on a neighboring O
    triangle according to the pattern in Fig.~\ref{fig:pattern}.
    Lengths are given in units of the Cu-Cu distance $a_0$.}
\label{tri}
\end{figure}

With a Cu-Cu distance $a_0 \approx 3.8 \rm{\AA{}}$, the side lengths
of the isosceles O-Cu-O triangle are $\frac{a_0}{2}$ for the two
equal legs and $\frac{a_0 }{\sqrt{2}}$ for the third side, as shown in
Fig.~\ref{tri}. The distance of the sides to the center-of-mass point
are given by $r_{\text{pd}}=\frac{a_0}{6}$ and
$r_{\text{pp}}=\frac{a_0}{6\sqrt{2}}$. Given the upper bound on the
current-current correlations $10^2\expect{\hat x|j_{k,k+\hat x}|\hat
  x}^2 \lesssim 0.1 (\frac{\rm{eV}}{\hbar})^2$, we infer a bound for the
particle current
\begin{equation}
j_{\text{pp}} < 0.03 \; \frac{\rm{eV}}{\hbar}.
\end{equation}
Note that the units correspond to $1 / \text{time}$ ($1 \rm{eV}/\hbar =
1.517 \times 10^{15}\; 1 / s$). Assuming that each triangle is
occupied by 1 hole only, this current corresponds classically to the
inverse time the hole takes to go around the triangle once.  

Assuming further that the hole dwells equal amounts of time on each link
defining the triangle, the velocity on the O-O link is given by
$$v_{\text{pp}}= \frac{a_0}{\sqrt{2}} /
\left(
  \frac{1}{3}\frac{1}{j_{\text{pp}}}\right)=\frac{3a_0}{\sqrt{2}}j_{\text{pp}}$$
and on the Cu-O links by 
$$v_{\text{pd}}=\frac{a_0}{2}/
\left(\frac{1}{3}\frac{1}{j_{\text{pp}}}\right)=\frac{3
  a_0}{2}j_{\text{pp}}.$$ With $L_z=m_e r_{\text{pp}}v_{\text{pp}}=m_e
r_{\text{pd}}v_{\text{pd}}$, we find an upper bound for the associated
angular momentum
%Hence, we
%get the electron (hole) velocity as $v=(1+1/\sqrt{2})a_0
%j_{\text{pp}}$, and by $L_z^{\text{pp}}=m_e r_{\text{pp}}
%v_{\text{pp}}$ we find the associated angular momentum
\begin{equation}
L_z^{\text{pp}}\lesssim 0.015\; \hbar .
\end{equation}
As there are two current carrying triangles per unit cell in the
pattern shown in Fig.~\ref{fig:pattern}, we find an upper bound of the
magnetic moment induced by the currents:
\begin{equation}
M_{\text{cell}}=2 M_{\bigtriangleup} \lesssim 0.03\; \mu_{\text{B}}.
\end{equation}
The result is below the estimate of $M\thickapprox 0.05 \text{--} 0.1
\mu_{\text{B}}$ found experimentally by Fauqu\'e
{\ea}~\cite{Fauque-06prl197001}.  (As we lower $\epsilon_\text{p}$ in
Section~\ref{sec:num} below, however, the upper bound for
$j_{\text{pp}}$ and $j_{\text{pd}}$ we are able to obtain from our
numerical experiments increase, and becomes comparable to the
experimental estimate.)

This provides an interesting perspective on the strength of the
magnetic moment observed in the experiment. If a pattern as shown in
Fig.~\ref{fig:pattern} were responsible for this moment, the
associated current would be in a range of $\sim 0.05 - 0.1 \; \rm{eV}
/\hbar$, a signal strength which we do not observe in the numerics.
Even if currents of this magnitude were present in the CuO layers,
however, following the previous derivation, the energy associated with
this current strength would only be in the range of $10 - 30 \;
\rm{meV}$ and thus too small to explain the phase diagram of the
high-$\text{T}_c$ cuprates.  This suggests that even if orbital
current alignment were responsible for the magnetic moment measured in
experiments, these currents would not be sufficiently large to account
for the strange metal phenomenology of high-$T_c$ superconductivity.

\begin{figure}[t]
%   \setlength{\unitlength}{2pt}
%   \begin{picture}(60,60)(0,0)
%   \multiput(20,57.7)(20,0){2}{\makebox(0,0){\small O}}
%   \multiput(0,37.7)(20,0) {4}{\makebox(0,0){\small O}}
%   \multiput(0,17.7)(20,0) {4}{\makebox(0,0){\small O}}
%   \multiput(20,-2.3)(20,0) {2}{\makebox(0,0){\small O}}
%   \multiput(30,47.7)(40,0){1}{\makebox(0,0){\small Cu}}
%   \multiput(10,27.7)(40,0){2}{\makebox(0,0){\small Cu}}
%   \multiput(30,7.7)(40,0){1}{\makebox(0,0){\small Cu}}
%   \multiput(30,57.3)(20,0){1}{\makebox(0,0){\vector(-1,0){14}}}
%   \multiput(10,37.3)(40,0){2}{\makebox(0,0){\vector(-1,0){14}}}
%   \multiput(30,40)(20,0){1}{\makebox(0,0){\vector(1,0){14}}}
%   \multiput(10,20)(40,0){2}{\makebox(0,0){\vector(1,0){14}}}
%   \multiput(30,17.3)(20,0){1}{\makebox(0,0){\vector(-1,0){14}}}
%   \multiput(30,0)(20,0){1}{\makebox(0,0){\vector(1,0){14}}}
%   \multiput(37,40)(20,-20){2}{{\vector(-1,1){5}}}
%   \multiput(17,20)(20,-20){2}{{\vector(-1,1){5}}}
%   \multiput(27,45)(-20,-20){2}{{\vector(-1,-1){5}}}
%   \multiput(47,25)(-20,-20){2}{{\vector(-1,-1){5}}} 
%   \multiput(22,55)( 20,-20){2}{{\vector(1,-1){5}}}
%   \multiput( 2,35)( 20,-20){2}{{\vector(1,-1){5}}}
%   \multiput(32,50)(-20,-20){2}{{\vector(1,1){5}}}
%   \multiput(52,30)(-20,-20){2}{{\vector(1,1){5}}}
%   \end{picture}
  \begin{minipage}[c]{0.4\linewidth}
    \includegraphics[width=\linewidth]{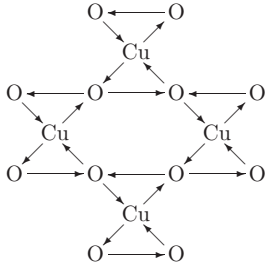}
  \end{minipage}
  \caption{An orbital current pattern proposed earlier by Varma,
    which has subsequently been ruled out by experiment.}
\label{fig:otherpattern}
\vspace{-4mm}
\end{figure}

The immediate conclusion we draw from our numerical results, however,
is that if the CuO planes are adequately described by a three-band
model with a set of couplings in the range we investigated, the
antiferromagnetic ordering observed by Fauqu\'e
{\ea}~\cite{Fauque-06prl197001} is not due to an orbital current
pattern as shown in Fig.~\ref{fig:pattern}.  We are hence led to
ponder whether our models might be consistent with an alternative
explanation of this experiment.  The correlations we observe would be
consistent with another orbital current pattern, shown in
Fig.~\ref{fig:otherpattern}, which had been proposed by Varma
earlier~\cite{Varma99prl3538, Varma06prb155113}.  This pattern,
however, has been successfully ruled out by neutron scattering
experiment~\cite{lee-99prb10405}, and is not consistent
with the experiment of Fauqu\'e {\ea}~\cite{Fauque-06prl197001}.  The
correlations we observe, on the other hand, provide no indication that
such a pattern is realized, as our system sizes are way to small to
establish the existence of any kind of long range order.  We are
merely not able to rule out order according to this pattern with our
numerical data.  Considering the possibility of this pattern being
realized, however, does in any event not bring us closer to
understanding the magnetic moment observed by Fauqu\'e
{\ea}~\cite{Fauque-06prl197001}.

\section{Spin-Spin correlations}
\label{sec:ss}

\begin{figure}[t]
  \begin{minipage}[c]{0.9\linewidth}
    \includegraphics[width=\linewidth]{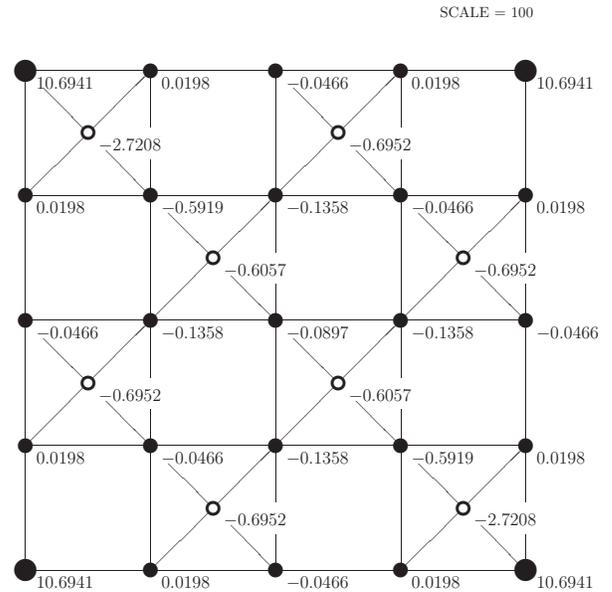}
  \end{minipage}
  \caption{Spin-spin correlations multiplied by $10^2$ for the ground
    state of \eqref{3btj} with $\epsilon_\text{p}=3.6$, $
    V_{\text{pd}}=1.2$ on a 24 site cluster (8 Cu = open circles, 16 O
    = filled circles).  Due to PBCs the O reference site is in all
    corners of the plot and indicated by a big black circle.}
\label{ssox}
\end{figure} 

If the observed magnetic moment is not due to orbital currents, what
other possibilities are there?  The only other explanation which comes
to mind within the confines of our three-band models is
antiferromagnetic order of the spins on the Oxygen lattice.  The
characteristic term driving the system into this kind of order would
be the antiferromagnetic coupling $J_{\text{pp}}$ in our effective
Hamiltonian~\eqref{3btj}.  According to our intuition about the
system, there is no reason to expect this kind of order, but as it is
straightforward to obtain the spin-spin correlations for our finite
cluster, we discuss them briefly in this section.

\begin{figure}[t]
  \begin{minipage}[c]{0.9\linewidth}
    \includegraphics[width=\linewidth]{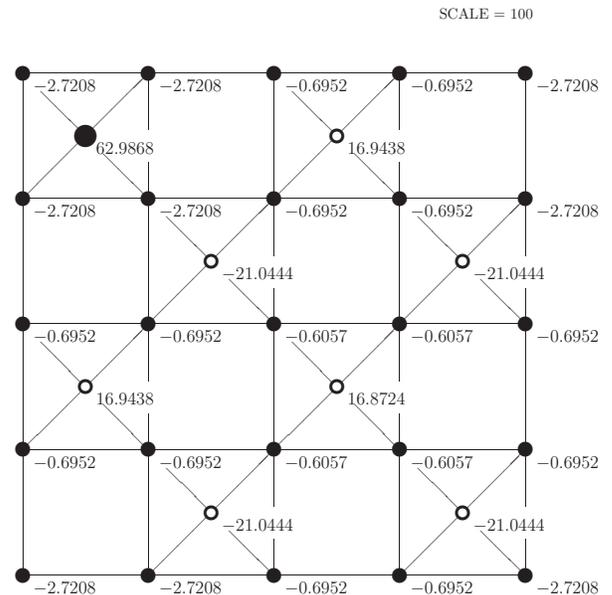}
  \end{minipage}
  \caption{Spin-spin correlations multiplied by $10^2$ for the ground
    state of \eqref{3btj} with $\epsilon_\text{p}=3.6$,
    $V_{\text{pd}}=1.2$ with PBCs.  The Cu reference site to the upper
    left is indicated by a big black circle.}
\label{sscu}
\end{figure}

We have evaluated the static spin-spin correlation of spins at sites
$i$ and $j$ given by $$\mathcal{S}_{ij}=\expect{\vec{S}_i\vec{S}_j},$$
with an O or a Cu as reference site.  The results are shown
in Figs.~\ref{ssox} and~\ref{sscu}, respectively.  For the O
sublattice, we only observe a very small staggered spin correlation
which falls off rapidly with distance on the scale of the lattice
constant, regardless of whether we choose an O or a Cu as
reference site.  We do not observe any indication of antiferromagnetic
order on the O sites, as the required long range correlations are
clearly absent.  We will hence not discuss the spin-spin correlations
any further and focus on the current-current correlations instead.

Regarding the experiment by Fauqu\'e
{\ea}~\cite{Fauque-06prl197001}, we believe that the explanation will
require a model which goes beyond the projected three-band models
studied here. The explanation might, for example, involve orbital
currents localized at the O atoms.

\section{Results for different parameter settings}
\label{sec:num}

We now consider various parameter choices for the three-band Hubbard
model. The two parameters which Varma~\cite{Varma06prb155113} assumed
to differ significantly from the values used above are the
relative on-site potential $\epsilon_{\text{p}}$, which he assumed to
be small compared to all other energy scales, and the inter-atomic
Coulomb interaction $V{_\text{pd}}$, which he assumed to be the
leading energy scale of the system generating the orbital currents.

Thus, our strategy is as follows. Firstly, we successively decrease
$\epsilon_{\text{p}}$ from 3.6 to 1.8, 0.9, 0.4, and finally 0, and
analyze each model as explained above. Secondly, for small values of
$\epsilon_{\text{p}}$, we double $V_{\text{pd}}$ from the standard
value 1.2 to 2.4 and look whether this significantly influences the
system. In doing so, we implicitly sweep over a broad
range of the charge transfer gap, which is dependent on
$V_{\text{pd}}$ and $\epsilon_{\text{pd}}$. To begin with, we have
computed the ground state energies per link for all inequivalent
points in the Brillouin zone for all parameter choices we consider
(see Table~\ref{tab:bz}).  We found that the ground states are either
situated in the $\Gamma$ point or M point of the Brillouin zone.  This
is not surprising as it is plausible that the ground state does not
carry any net momentum, and also consistent with the current pattern
shown in Fig.~\ref{fig:pattern}. For the states at the $\Gamma$ and the
M point, we then implement the rotation symmetry discussed above and
compute the ground state energies per link in the respective subspaces
exactly. The results are shown in Table~\ref{tab:bzf}.

\begin{table}[h]
  \centering
   \begin{tabular}{cc|cccccc}
     \hline\hline
     $\epsilon_{\text{p}}$   & $V_{\text{pd}}$ &  \phantom{oooo}(0,0)\phantom{oo} 
     & \phantom{oo}($\frac{\pi}{2}$,0)\phantom{oo} 
     &\phantom{oo} ($\pi$,0)\phantom{oo} &\phantom{oo} (0,$\pi$)\phantom{oo}  \\[0.5ex]
     \hline
     3.6  &  1.2   &  -0.8473  &   -0.8356  &   \underline{-0.8520}  &   -0.8359  \\
     1.8  &  1.2   & \underline{ -1.4187}  &   -1.3888  &   -1.4104  &   -1.3925  \\
     0.9  &  1.2   & \underline{ -1.7852}  &   -1.7508  &   -1.7732  &   -1.7569  \\
     0.4  &  1.2   & \underline{ -2.0290}  &   -2.0011  &   -2.0185  &   -2.0066  \\
     0.4  &  2.4   & \underline{ -1.6486}  &   -1.6381  &   -1.6433  &   -1.6387  \\
     0.0  &  1.2   & \underline{ -2.2543}  &   -2.2427  &   -2.2438  &   -2.2407  \\
     0.0  &  2.4   &  -1.9199  &   -1.9265  &   \underline{-1.9342}  &   -1.9261  \\
     \hline\hline
   \end{tabular}
   \caption{Ground state energies per unit cell for the different momentum 
subspaces in the $N_{\text{link}}^{\text{max}}=4$ approximation.  The numbers for the subspace with the 
lowest energies, corresponding to the ground state of the full Hamiltonian, are underlined.}
   \label{tab:bz}
  \end{table}

\begin{widetext} \vspace{-2pt}

    \centering
  \begin{table}[h]
    \centering
   \begin{tabular}{cc|ccc|ccc}
     \hline\hline
     $\epsilon_{\text{p}}$   & $V_{\text{pd}}$ &  \phantom{ooo}(0,0,0)\phantom{oo} 
     & \phantom{oo}(0,0,$\frac{\pi}{2}$)\phantom{oo} &\phantom{oo} (0,0,$\pi$)\phantom{oo} 
     &\phantom{oo} ($\pi$,0,0)\phantom{oo} &\phantom{oo} ($\pi$,0,$\frac{\pi}{2}$)\phantom{oo} 
     &\phantom{oo} ($\pi$,0,$\pi$)\phantom{oo} \\[0.5ex]
     \hline
     3.6  &  1.2   &  -0.8544  &   -0.8545  &   -0.8843  &  -0.8513  &   -0.8570  &   \underline{-0.8883}   \\
     1.8  &  1.2   &  -1.4551  &   -1.4350  &   \underline{-1.4857}  &   -1.4462  &   -1.4448  &   -1.4769   \\
     0.9  &  1.2   &  -1.8440  &   -1.8136  &   \underline{-1.8672}  &   -1.8345  &   -1.8304  &   -1.8557   \\
     0.4  &  1.2   &  -2.0986  &   -2.0820  &   \underline{-2.1149}  &   -2.0932  &   -2.0859  &   -2.1053   \\
     0.4  &  2.4   &  -1.6864  &   -1.6788  &   \underline{-1.7007}  &   -1.6950  &   -1.6741  &   -1.6923   \\
     0.0  &  1.2   &  -2.3269  &   \underline{-2.3400}  &   -2.3389  &   -2.3317  &   -2.3280  &   -2.3372   \\
     0.0  &  2.4   &  -1.9508  &   -1.9613  &   -1.9647  &   \underline{-1.9766}  &   -1.9543  &   -1.9619   \\
     \hline\hline
   \end{tabular}

   \caption{Exact ground state energies for the $\Gamma$ point 
     $(0,0,q_{\text{rot}})$, and the M point $(\pi,0,q_{\text{rot}})$, 
     with additionally applied rotation symmetry ($q_{\text{rot}}=3\pi/2$ 
     is degenerate to $q_{\text{rot}}=\pi/2$).  The numbers for the subspace 
     with the lowest energies, corresponding to the ground state of the 
     full Hamiltonian, are underlined.}
   \label{tab:bzf}
  \end{table}
\end{widetext}

We now turn to the different choices for
the model parameters.  Except for $\epsilon_{\text{p}}$ and
$V{_\text{pd}}$, we use the three-band Hubbard parameters calculated
by Hybertsen \ea~\cite{Hybertsen-89prb9028}. We label the different
sections by the values for $\epsilon_{\text{p}}$ and $V{_\text{pd}}$
we specifically investigate.

\subsection{$\epsilon_{\text{p}}=1.8$, $V{_\text{pd}}=1.2$}
\begin{figure}[t]
  \begin{minipage}[c]{0.9\linewidth}
    \includegraphics[width=\linewidth]{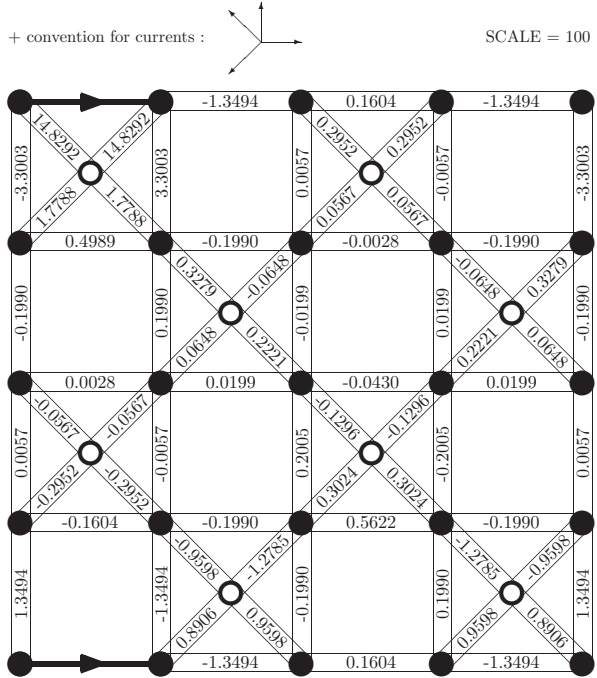}
  \end{minipage}
  \caption{Current-current correlations $\expect{j_{k,k+\hat
        x}j_{l,m}}$ multiplied by $10^2$ for the ground state of
    \eqref{3btj} with $\epsilon_\text{p}=1.8, \; V_{\text{pd}}=1.2$ on
    a 24 site cluster with PBCs (reference link indicated in the top
    and bottom left corner).}
\label{fig:ep18}
\vspace{-3mm}
\end{figure} 
As we decrease $\epsilon_{\text{p}}$ from 3.6 to 1.8, the ground state
switches from momentum $(\pi,0)$ to $(0,0)$, \ie from the M to the
$\Gamma$ point in the Brillouin zone (see Table~\ref{tab:bz}).
Implementing the rotational symmetry yields $E_{(0,0,0)}=-1.4551$,
$E_{(0,0,\pi/2)}=-1.4350$, and $E_{(0,0,\pi)}=-1.4857$ (see
Table~\ref{tab:bzf}), \ie the ground state is at $(0,0,\pi)$. The
current-current correlations for this state are depicted in
Fig.~\ref{fig:ep18}. There is no evidence for a pattern along the
lines of Fig.~\ref{fig:pattern}. Instead, the correlations decrease
rapidly with distance.  The ``fluctuations'' or ``noise'' inherent in
the finite size calculation are comparable to the preceding case
$\epsilon_{\text{p}}=3.6$. The values of
$10^2\!\cdot\!\expect{j_{k,k+\hat x}j_{l,l+\hat x}}$ for the four
horizontally connected links in the center of Fig.~\ref{fig:ep18},
$+0.0028, +0.0199, -0.0430$, and $+0.0199$ indicate no relevant scale
of correlations. Recall that the orbital current pattern shown in
Fig.~\ref{fig:pattern} would require all the numbers to be positive.

\subsection{$\epsilon_{\text{p}}=0.9$, $V{_\text{pd}}=1.2$}
\begin{figure}[t]
  \begin{minipage}[c]{0.9\linewidth}
    \includegraphics[width=\linewidth]{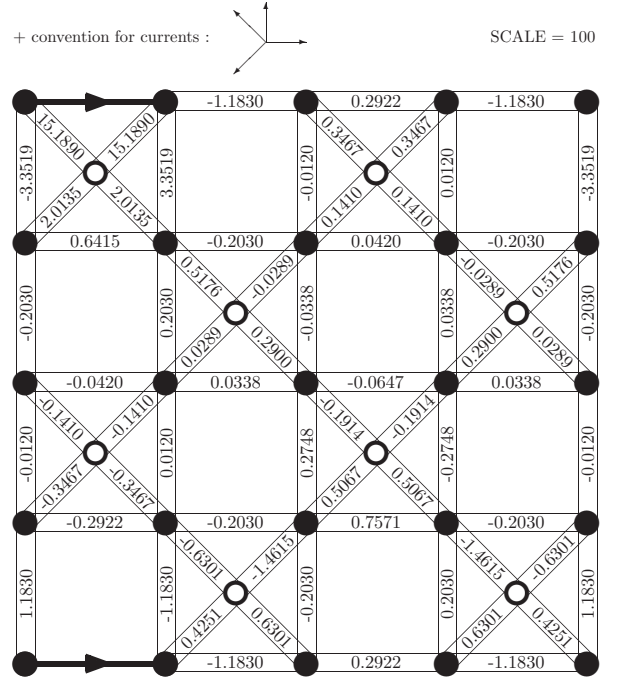}
  \end{minipage}
  \caption{Current-current correlations $\expect{j_{k,k+\hat
        x}j_{l,m}}$ multiplied by $10^2$ for the ground state of
    \eqref{3btj} with $\epsilon_\text{p}=0.9$, $V_{\text{pd}}=1.2$ on
    a 24 site cluster with PBCs.}
\label{fig:ep09}
%\vspace{-3mm}
\end{figure} 
As we decrease $\epsilon_{\text{p}}$ further to 0.9, the lowest state
remains in the $\Gamma$ point, \ie $(0,0,\pi)$.  The ground state
correlations are depicted in Fig.~\ref{fig:ep09}.  They fall off less
rapidly with distance than for larger values of $\epsilon_{\text{p}}$. 
Again, there is no indication of a current pattern.  The values of
$10^2\!\cdot\!\expect{j_{k,k+\hat x}j_{l,l+\hat x}}$ for the four
horizontally connected links in the center of Fig.~\ref{fig:ep09},
$-0.420, +0.0338, -0.0647$, and $+0.0338$, indicate likewise no relevant scale
of correlations.

\subsection{$\epsilon_{\text{p}}=0.4$, $V{_\text{pd}}=1.2$}
\begin{figure}[h]
  \begin{minipage}[c]{0.9\linewidth}
    \includegraphics[width=\linewidth]{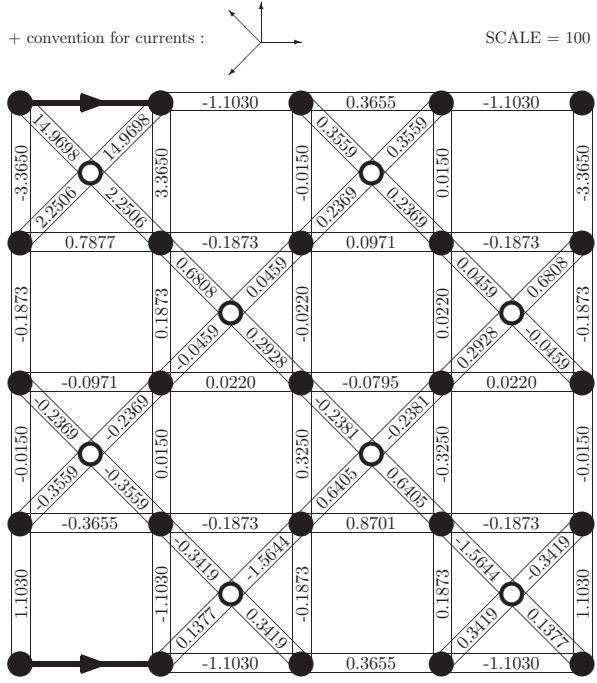}
  \end{minipage}
  \caption{Current-current correlations $\expect{j_{k,k+\hat
        x}j_{l,m}}$ multiplied by $10^2$ for the ground state of
    \eqref{3btj} with $\epsilon_\text{p}=0.4$, $V_{\text{pd}}=1.2$ on a 24 site
    cluster with PBCs.}
\label{fig:ep04}
%\vspace{-3mm}
\end{figure}
The correlations for the ground state, which remains at the
$(0,0,\pi)$ point, are depicted in Fig.~\ref{fig:ep04}.  We have now
reached a parameter regime for which Varma proposed that a current
pattern should occur: $\epsilon_{\text{p}}$ is small compared to
$V_{\text{pd}}$, and $V_{\text{pd}}$ is of the order of the other
scales.  However, we still find no indication of a current pattern.
The values $10^2\!\cdot\!\expect{j_{k,k+\hat x}j_{l,l+\hat x}}$ for
the four horizontally connected links in the center of
Fig.~\ref{fig:ep04}, $-0.0971, +0.0220, -0.0795$, and $+0.0220$,
remain small.  Nonetheless, let us estimate the energy associated with
an upper bound for the currents as elaborated on above.  We estimate
$10^2\!\cdot\!\expect{j_{k,k+\hat x}}^2 < 0.1$ and hence
$\expect{\hat x|j_{k,k+\hat x}|\hat x}^2 < 2\cdot 10^{-3}$ as an upper
bound for a uniform contribution according to a current pattern as
depicted in Fig.~\ref{fig:pattern}.  As above, $\expect{\hat
  x|j_{k,k+\hat x}|\hat x}$ is denoted by $j_\text{pp}$.  For the
kinetic energy $\varepsilon_\text{pp}$ per link associated with a
spontaneous current 
\begin{equation}\nonumber
j_{\text{pp}} \lesssim 0.05 \; \frac{\rm{eV}}{\hbar}
\end{equation}
we obtain with $n_\text{p}=0.24$
\begin{equation}
  \varepsilon_\text{pp}
  \approx\frac{\hbar^2 j_\text{pp}^2}{4t_\text{pp}n_\text{p}}<3\cdot 10^{-3}.
  \nonumber
\end{equation}
If a spontaneous currents were hence to exist, the energy associated
with them would be too small to allow for an interpretation of the
strange metal phase in terms of the quantum critical fluctuations
around this phase.

\subsection{$\epsilon_{\text{p}}=0.4$, $V{_\text{pd}}=2.4$}
\begin{figure}[h]
  \begin{minipage}[c]{0.9\linewidth}
    \includegraphics[width=\linewidth]{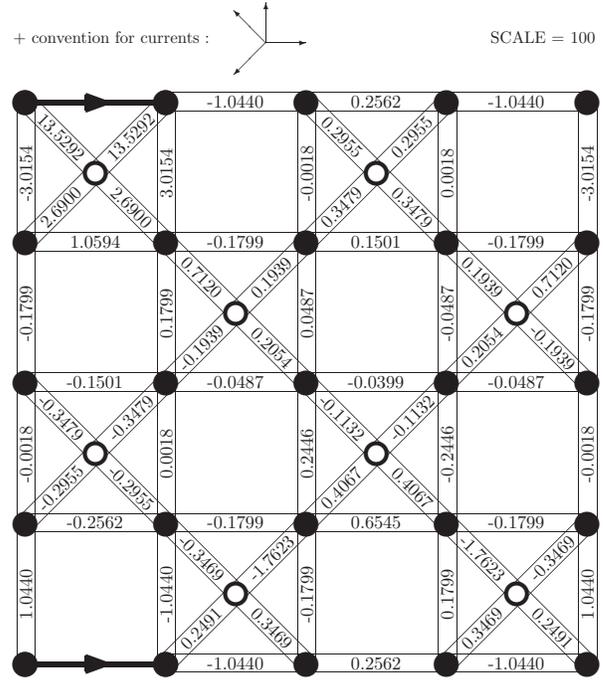}
  \end{minipage}
  \caption{Current-current correlations $\expect{j_{k,k+\hat
        x}j_{l,m}}$ multiplied by $10^2$ for the ground state of
    \eqref{3btj} with $\epsilon_\text{p}=0.4$, $V_{\text{pd}}=2.4$ on a 24 site 
    cluster with PBCs.}
\label{fig:ep04v24}
%\vspace{-3mm}
\end{figure} 
For $\epsilon_{\text{p}}=0.4$, we have also doubled the Coulomb
repulsion between the Cu and O sites to $V{_\text{pd}}=2.4$, which
then becomes the leading energy scale before the Copper-Oxygen hopping
$t_{\text{pd}}=1.5$.  The ground state remains at the $\Gamma$ point
of the Brillouin zone (see Table~\ref{tab:bz}). Compared to
$V{_\text{pd}}=1.2$, the energies of the $\Gamma$ and M ground states
are now much closer to each other.  The correlations for the ground
state at $(0,0,\pi)$ are shown in Fig.~\ref{fig:ep04v24}.  Again,
there is no evidence of a current pattern. The values of
$10^2\!\cdot\!\expect{j_{k,k+\hat x}j_{l,l+\hat x}}$ for the four
horizontally connected links in the center of Fig.~\ref{fig:ep04v24},
$-0.1501, -0.0487, -0.0399$, and $-0.0487$, are comparable to the case
$V_{\text{pd}}=1.2$ discussed above. Note that the four horizontal
links we consider are now aligned. Unfortunately, the sign of all
four numbers is opposite to the sign required by the pattern
shown in Fig.~\ref{fig:pattern}.

\subsection{$\epsilon_{\text{p}}=0$, $V{_\text{pd}}=1.2$}
\begin{figure}[h]
  \begin{minipage}[c]{0.9\linewidth}
    \includegraphics[width=\linewidth]{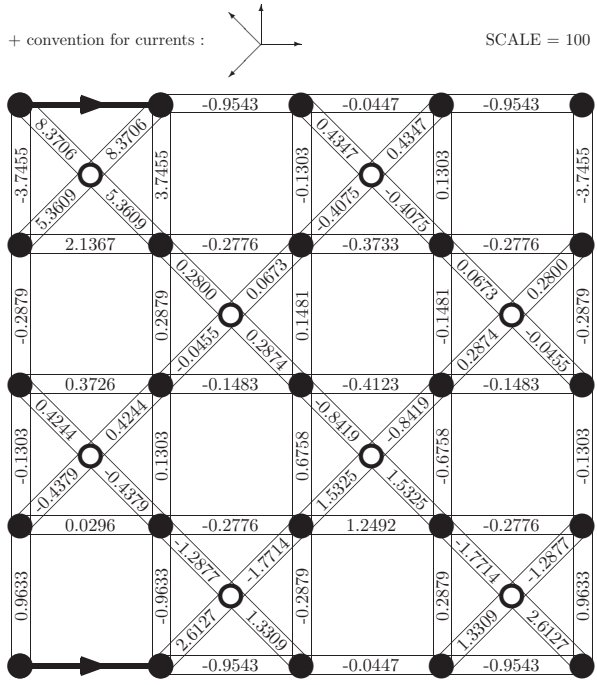}
  \end{minipage}
  \caption{Current-current correlations $\expect{j_{k,k+\hat
        x}j_{l,m}}$ multiplied by $10^2$ for the ground state of
    \eqref{3btj} with $\epsilon_\text{p}=0.0$, $V_\text{pd}=1.2$ on a 24 site cluster  with PBCs.}
\label{fig:ep00}
%\vspace{-3mm}
\end{figure}
\begin{figure}[h]
  \begin{minipage}[c]{0.9\linewidth}
    \includegraphics[width=\linewidth]{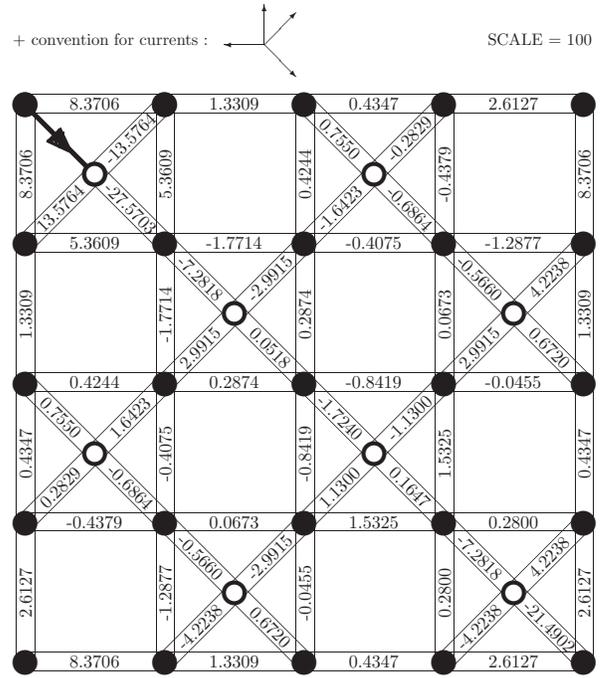}
  \end{minipage}
  \caption{Current-current correlations $\expect{j_{k,k+\hat
        x}j_{l,m}}$ multiplied by $10^2$ for the ground state of
    \eqref{3btj} with $\epsilon_\text{p}=0.0$, $V_\text{pd}=1.2$ on a
    24 site cluster with PBCs. The Cu-O reference link is indicated by
    a black arrow in the upper left corner.}
\label{fig:ep00-2}
\vspace{-3mm}
\end{figure}
This setting has already been discussed
previously~\cite{Greiter-07prl027005}. In a sense,
$\epsilon_{\text{p}}=0$ is the most favorable choice for Varma's mean
field approach. It should be kept in mind, however, that
$\epsilon_{\text{p}}$ must be positive and finite in the experimental
system to account for the antiferromagnetic order in the undoped
cuprates.  The ground state is now doubly degenerate and found at
$(0,0,\pm \pi/2)$.  There is no current pattern observable, but the
values of $10^2\!\cdot\!\expect{j_{k,k+\hat x}j_{l,l+\hat x}}$ for the
four horizontally connected links in the center of
Fig.~\ref{fig:ep00}, $0.3726, -0.1483, -0.4123$, and $-0.1483$, are
larger than for any of the other parameter settings we investigated.
The uniform current is still zero, but with larger ``fluctuations'' or
``noise'' due to the finite size of our system.  We estimate an upper
bound for a uniform positive correlation
$10^2\!\cdot\!\expect{j_{k,k+\hat x}}^2 < 0.2$ and hence $\expect{\hat
  x|j_{k,k+\hat x}|\hat x}^2 < 4\cdot 10^{-3}$.  As above,
$\expect{\hat x|j_{k,k+\hat x}|\hat x}$ is denoted by
$j_\text{pp}$, %_\text{O-O}$.
and the approximate kinetic energy $\varepsilon_\text{pp}$ per link
associated with a spontaneous current 
\begin{equation}\nonumber
  j_{\text{pp}} < 0.06 \; \frac{\rm{eV}}{\hbar}
\end{equation}
 yields with
\eqref{eq:epp} and $n_{\text{p}}=0.34$,
\begin{equation}
  \varepsilon_\text{pp}<5\cdot 10^{-3}.
  \nonumber
\end{equation}
The correlations with Cu-O reference link are shown in
Fig.~\ref{fig:ep00-2}.  A similar analysis yields with \eqref{eq:epd}
\begin{equation}
  \varepsilon_\text{pd} <4 \cdot 10^{-3}.
  \nonumber
\end{equation}
The upper bound of 5 meV we found for the energy associated with
spontaneous currents corresponds to a temperature
$T_{\text{current}}\sim 60 \text{K}$.  To explain the strange metal
phase, however, the energy scale responsible for the quantum critical
fluctuations would have to extend at least to a range of several
hundred Kelvin.

Note that the upper bound for the magnetic moment per unit cell is now
given by 
\begin{equation}
M_{\text{cell}}=2 M_{\bigtriangleup} \lesssim 0.06\; \mu_{\text{B}},
\end{equation}
a value roughly comparable to the range of $M\thickapprox 0.05
\text{--} 0.1 \mu_{\text{B}}$ found experimentally by Fauqu\'e
{\ea}~\cite{Fauque-06prl197001}.  This does not imply that these
orbital currents exist, but only states that if we assume
a three band model with
$\epsilon_{\text{p}}=0$, we are not able to rule out an orbital
current pattern as shown in Fig.~\ref{fig:pattern} as an explanation
for the experimentally observed magnetic moment.  On the other hand,
even if the observed moments are due to such a pattern, the energies
involved are too small to explain the phenomenology of the strange
metal phase.

\subsection{$\epsilon_{\text{p}}=0$, $V{_\text{pd}}=2.4$}
\begin{figure}[t]
  \begin{minipage}[c]{0.9\linewidth}
    \includegraphics[width=\linewidth]{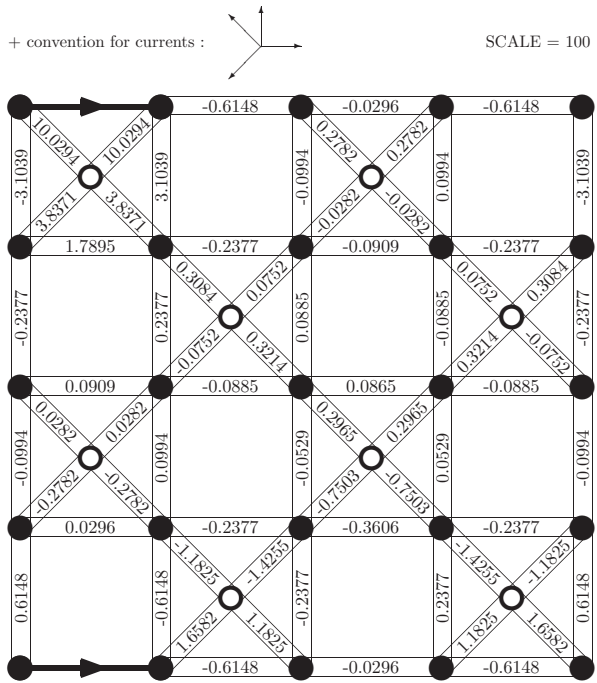}
  \end{minipage}
  \caption{Current-current correlations $\expect{j_{k,k+\hat
        x}j_{l,m}}$ multiplied by $10^2$ for the ground state of
    \eqref{3btj} with $\epsilon_\text{p}=0.0$, $V_\text{pd}=2.4$ on a 24 site 
    cluster with PBCs.}
\label{fig:ep00v24}
%\vspace{-3mm}
\end{figure}
Finally, we double the repulsion $V_{\text{pd}}$ for $\epsilon_\text{p}=0$.
The ground state is now at the M point, at $(\pi,0,0)$ (see
Tab.~\ref{tab:bzf}).  The corresponding correlations are shown in
Fig.~\ref{fig:ep00v24}.  Again, there is no evidence for a current
pattern.  The values of $10^2\!\cdot\!\expect{j_{k,k+\hat
    x}j_{l,l+\hat x}}$ for the four horizontally connected links in
the center of Fig.~\ref{fig:ep00v24}, $0.0909, -0.0885, 0.0865$, and
$-0.0885$, are rather small.  We estimate an upper bound
$10^2\!\cdot\!\expect{j_{k,k+\hat x}}^2 < 0.1$ and hence $\expect{\hat
  x|j_{k,k+\hat x}|\hat x}^2 < 2\cdot 10^{-3}$ for the contribution
from spontaneous currents.  For the associated kinetic energy
$\varepsilon_\text{pp}$ per link we find with $n_{\text{p}}=0.36$
% \begin{equation}
%   \label{en}
%   J^\text{pp}=n_\text{p}v,\quad 
%   \varepsilon^\text{pp}=\frac{1}{2}n_\text{p} m v^2,\quad 
%   m=\frac{1}{2t_\text{pp}},
%   \nonumber
% \end{equation}
\begin{equation}
  \varepsilon_\text{pp}<2\cdot 10^{-3}.\nonumber
\end{equation}
Note that since $V_{\text{pd}}$ is driving the orbital currents in the
mean field theory proposed by Varma~\cite{Varma06prb155113}, his
theory would predict the model to show the strongest propensity to
form current patterns for this choice of parameters.  By contrast, we
find no indication of such a propensity in our numerical experiments.

\section{Finite size effects}
\label{sec:finite}

The numerical calculations we report here were performed on a
cluster of 8 unit cells, \ie 24 sites (8 Copper and 16 Oxygen).  The
question we wish to address in this section is whether any of the
finite size effects might affect our overall conclusion.  Due to the
finite size, we have not been able to measure the current-current
correlations at long distances, where they would accurately provide
the size of a spontaneous current if such a current were to exist.
Instead, we have only been able to establish upper bounds on such
currents.  These upper bounds, however, turned out to be small enough
to allow us to rule out that an orbital current pattern as proposed by
Varma and shown in Fig.~\ref{fig:pattern} is responsible for the
anomalous properties of the strange metal phase in the cuprates.

There are, however, two other finite size effects which may limit the
validity of our conclusion.  The first is that the ground states in
our systems have spin $S=1/2$, as we have a total of 9 holes,
corresponding to a doping of one hole away from half filling.  The
current carrying state proposed and investigated by Varma, by
contrast, is  in general a spin singlet.  Could it be that the extra
spin 1/2 destroys the orbital current pattern in our numerical
experiments, while the currents would be present if we had an
infinite system?  We will argue now that any possible effect would not
affect our conclusions.

To begin with, assuming that no magnetic spin order is present, the
ground state of the infinite system will be a spin singlet, regardless
of whether the state carries a spontaneous current or not.  This also
holds for any finite system with an even number of electrons, even if
long range antiferromagnetic correlations in the spins were present.
The ground state for a finite system with an odd number of electrons,
as we have investigated in this work, will correspond to the ground
state for an even number of electrons supplemented by an excitation
which carries spin 1/2.  This excitation will cost a finite amount of
energy.  The question relevant for the validity of our conclusion is
whether the energy cost of this spin 1/2 excitation is higher for a
current carrying state than it is for a state without currents, and if
it is higher, by which amount.  Since the term driving the
spontaneous currents in Varma's analysis is $V_{\text{pd}}$, the
energy associated with this term has to be lower in the current carrying
state, while we would expect most other terms in the Hamiltonian, but
in particular the antiferromagnetic exchange terms $J_{\text{pd}}$ and
$J_{\text{pp}}$, to be slightly higher in energy.  As the excitation
energy for the extra spin 1/2 amounts mostly to an extra energy cost
in $J_{\text{pd}}$ and $J_{\text{pp}}$, we expect that this excitation
will cost less energy in the current carrying state than in the state
without currents.  In other words, unless the antiferromagnetic
exchange terms in~\eqref{3bH} were to contribute towards driving the
system into a current carrying phase, the spin 1/2 excitation in our
finite system would enhance the stability of this phase.  The fact
that we do not observe a current pattern in the presence of the extra
spin 1/2 makes our conclusion even more robust than it would be
without the excitation.

To rule out any doubt completely, let us be unreasonable and assume
that the antiferromagnetic exchange terms do in fact enhance the
systems propensity to develop spontaneous currents, and that the
antiferromagnetic exchange energy in the current carrying phase is
maybe about 10\% below the energy of the phase without the currents.
For the parameter choice $\epsilon_{\text{p}}=0$ and
$V_{\text{pd}}=1.2$, the antiferromagnetic exchange energy per unit
cell is
\begin{equation}
\epsilon_{\text{J}}
=\frac{1}{8}\bra{\Psi_0}H_{\text{J}}\ket{\Psi_0}=122 \; \text{meV},
\end{equation}
with $H_{\text{J}}$ given in~\eqref{3btj}.  To estimate the energy cost
of the spin 1/2 excitation, we compare the ground state energy for
$S=1/2$ obtained for 5 up and 4 down spin holes with the ground state
energy for $S=3/2$ obtained for 6 up and 3 down spin holes, and find
an energy difference of $\Delta=230 \; \text{meV}$.  Since the ground
state energy for the different spin sectors should roughly be
proportional to $S^2$, the energy cost of the $S=1/2$ excitation
should be roughly 1/8th of this difference, or about 30 meV.  The
energy cost of the excitation per unit cell of our finite cluster is
hence of the order of 4 meV.  If we now assume that the energy cost
for this $S=1/2$ excitation increases by 10\% if spontaneous currents
are present (the cost increases because the excitation disturbs the
current driving antiferromagnetic correlations), the additional energy
cost for the currents due to the extra spin 1/2 would be of order 0.4
meV per unit cell, or roughly 0.04 meV per link.  If this energy cost
were to destabilize the spontaneous currents, the energy associated
with them would be below the upper bounds estimated from the
current-current correlations above.

The second finite size effect we briefly wish to mention is that the
special geometry of periodic boundary conditions might have an
unintended influence on the system, maybe in that it stabilizes a
state without currents which would not be stable in the infinite
system.  We have hence twisted and varied the boundary conditions in
any way we could think of, but found that the correlations, and in
particular the ``fluctuation'' or ``noise level'' due to the finite
size which limits or ability to put upper bounds on the correlations,
remained unchanged.

\section{Conclusion}
\label{sec:con}
Let us summarize the results of our numerical studies on three-band
Hubbard models for the cuprate planes in CuO superconductors. For the
commonly accepted parameter values as calculated by
Hybertsen~\cite{Hybertsen-89prb9028}, we find no orbital current
pattern as shown in Fig.~\ref{fig:pattern} as well as no significant
antiferromagnetic spin-spin correlation on the Oxygen sites. As we
sweep over a considerable parameter regime of $\epsilon_{\text{p}}$
and $V_{\text{pd}}$ and compute the current-current correlations for
the respective ground states, we likewise do not observe any evidence
for a pattern as advocated by Varma's mean field analysis.  Instead,
we find that the correlations change quantitatively, but not
qualitatively, as we move through the parameter space. We also observe
that there is no clear dependence of the correlations on
$V_{\text{pd}}$, which is not consistent with what one would expect
from Varma's analysis. We conclude that while we cannot rule out that
orbital current patterns exist, we can rule out that they are
responsible for the properties of the strange metal phase or the
anomalous normal state properties extending up to temperatures of
several hundred Kelvin, as the energy associated with the spontaneous
loop currents would be less than 5 meV per link if such currents were
to exist.  We have assumed that the CuO planes are adequately
described by the three-band Hubbard model \eqref{3bH}, but we have
allowed $\epsilon_{\text{p}}$ to be much smaller and $V_{\text{pd}}$
larger than generally agreed upon, and based our estimate for the
upper bound of 5 meV on the for our purposes most unfavorable case
$\epsilon_{\text{p}}=0$.

%This result calls the validity of Varma's mean field analysis in question.

\begin{acknowledgments}
  We wish like to thank C.M.\ Varma, V.\ Aji, F.\ Evers, and in
  particular P.\ W\"olfle for many illuminating discussions of this
  subject.  RT was supported by a PhD scholarship from the
  Studienstiftung des deutschen Volkes.  We further acknowledge the
  support of the computing facilities of the INT at the
  Forschungszentrum Karlsruhe.
\end{acknowledgments}

\vspace{10pt}
{\it Note added:} After this work was completed, we learned of a slave
boson mean field calculation by Kremer, Brinckmann, and
W\"olfle~\cite{Kremer-07}, who likewise found no spontaneous currents
in the ground state of the model originally analyzed by Varma.

% Create the reference section using BibTeX:
%\bibliographystyle{/users/tkm/rachel/bib/prsty}
%\bibliography{/users/tkm/rachel/bib/book,/users/tkm/rachel/bib/paper,/users/tkm/rachel/bib/unpub,/users/tkm/rachel/bib/htc}

\end{document}